# TERAHERTZ AND INFRARED UNCOOLED DETECTOR BASED ON A MICROCANTILEVER AS A RADIATION PRESSURE SENSOR

Gennady P. Berman and Boris M. Chernobrod

*Theoretical Division, Los Alamos National Laboratory MS 213, Los Alamos, New Mexico 87545*

Alan R. Bishop

*Theory, Simulation & Computation Directorate, MS B210, Los Alamos National Laboratory, Los Alamos, New Mexico 87545*

Vyacheslav N. Gorshkov

*Theoretical Division & CNLS, MS B213, Los Alamos National Laboratory, Los Alamos, New Mexico 87545, & The Institute of Physics, National Academy of Sciences of Ukraine, Nauki Ave. 46, Kiev-39, 03650, MSP-65, Ukraine*

## Abstract

We consider a far infrared (terahertz), room-temperature detector based on a microcantilever sensor of the radiation pressure. This system has a significantly higher sensitivity than existing uncooled detectors in the far infrared (terahertz) spectral region. The significant enhancement of sensitivity is due the combination non-absorption detection method and high quality optical microcavity. Our theoretical analysis of the detector sensitivity and numerical simulations demonstrate that the narrowband heterodyne detector with the band width 30 MHz has a minimal measurable intensity by three orders of magnitude less than conventional uncooled detectors. In the case of the broadband detector, the noise equivalent temperature difference (NETD) is 7.6 mK, which is significantly smaller than for conventional uncooled thermal detectors.

## I. Introduction

Improvement of the sensitivity of detection in the far infrared (terahertz) region is very important for many applications including remote sensing of explosive materials, chemical, and biological agents, surveillance, night-vision, and medical imaging. Uncooled thermal detectors are very desirable because cooling systems add cost and cause reliability problems that are incompatible with most applications. Recently, significant progress has been demonstrated for uncooled thermal detectors based on microcantilever arrays[1-4]. Microcantilever systems are potentially more sensitive and have shorter response times than conventional thermoelectric and semiconductor solid state detectors. In conventional thermoelectric imagers the temperature rise in each pixel is measured electrically. The electrical connectivity to each pixel causes prohibitive complexity and cost. A microcantilever-based imager could have full optical readout, eliminating readout electronics, a very attractive feature. Currently microcantilever-based detectors exploit the thermo-mechanical effect, in which a bilayer microcantilever illuminated by radiation exhibits banding due to the difference in the thermal expansion coefficients of the different layers. This measurement method has some limitations for the sensitivity and response time. In the present paper, we propose to utilize the radiation pressure to measure the radiation power. We showed recently[5,6] that, using the radiation pressure, the microcantilever detector significantly improves the sensitivity of the frequency modulation spectroscopy. One of the significant advantages of radiation pressure measurements is the possibility of using a high quality microcavity, which leads to a significant sensitivity enhancement due to this non-absorption mechanism of detection. Note that application of an optical microcavity to absorption detectors leads to a moderate enhancement of sensitivity because absorption at the photosensitive surface affects the quality of optical resonator. We consider two applications: i) a narrowband heterodyne detector and ii) a broadband detector. The narrowband heterodyne detector could be used as a miniature spectrometer. The broadband detector potentially could be used in a new type of thermal imager. In the case of the narrowband detector, a microcantilever senses the beats between the heterodyne field and spectral components of the signal. The frequencies of the signal spectral components which we detect are shifted to the red and blue relative to the heterodyne frequency by an amount equal to the cantilever resonance frequency. The spectrum of the signal can be scanned by tuning the heterodyne frequency.

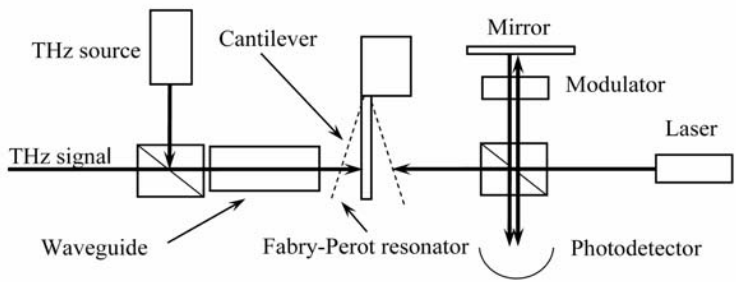

Fig. 1. Setup of the microcantilever based narrowband heterodyne detector.

Fortunately, compact, uncooled far infrared solid state lasers are available[7,8]. Application of the standard microcavity techniques gives an additional increase of sensitivity due to cavity enhancements of both the heterodyne and the signal fields. In the present paper, we describe the results of our theoretical analysis and numerical simulations of the microcantilever-based detectors schematically depicted in Figs. 1 and 2. We show that i) the proposed narrowband detector is expected to have a sensitivity three orders of magnitude better than the typical sensitivity of conventional uncooled far infrared (terahertz) detectors, and ii) the broadband detector potentially has the sensitivity one order of magnitude better than conventional uncooled thermal detectors.

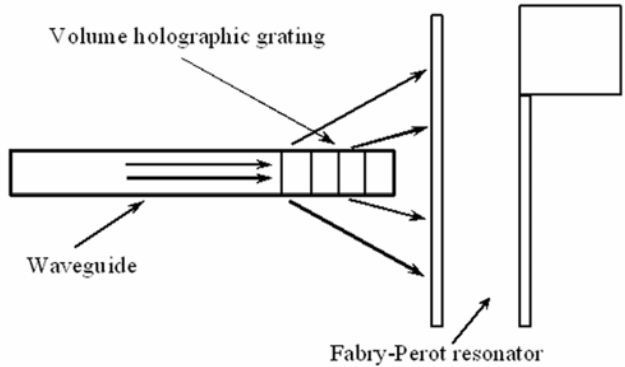

Fig. 2. Setup of the broadband thermal radiation detector.

## II. Sensitivity analysis of a narrow band heterodyne detector

Figure 1 shows a possible setup for a microcantilever-based narrowband detector as a radiation pressure sensor. This optical scheme is similar to that used for efficient laser cooling of microcantilever[9,10], which could cool the microcantilever to its quantum mechanical ground state. The radiation of the heterodyne laser is mixed with the signal and sent through an optical waveguide, the other end of which is polished and coated with a high reflectivity material (for example a Bragg mirror).

The coated waveguide end, in combination with the coated surface of the cantilever, forms a Fabry-Perot optical resonator. The cantilever oscillations are measured by a Michelson interferometer. We use a model that describes the oscillations of the cantilever as a damped harmonic oscillator driven by light pressure and thermal noise. The fields in the Fabry-Perot cavity are described by resonator equations, which include damping and incident waves. The electromagnetic fields are written in the form

$$E = \frac{1}{2}\left(E_h(t)\exp(-i\omega_h t) + E_s(t)\exp(-i\omega_s t)\right) + c.c., \qquad (1)$$

where $E_{h,s}(t)$, $\omega_{h,s}$ are the amplitude and the frequency of the heterodyne and signal fields, correspondingly. The slow field amplitudes inside the resonator satisfy the following equations:

$$\frac{dE_{h,s}}{dt} = -\frac{1}{\tau_p}\left(E_{h,s} - E_{h,s}^0(t)\right), \qquad (2)$$

where the damping time of the resonator is $2\tau_p^{-1} = (1-R)c/(2L\sqrt{R})$, $R = R_1 R_2$, $R_{1,2}$ are the reflection coefficients of the fiber end and the microcantilever surface, correspondingly, and L is the average resonator length. In the case of the steady-state field amplitudes, $E_{h,s}^0$, are

$$E_{h,s}^0 = \frac{E_{h,s}^{ext} T_0}{1 - R\exp(\delta_{h,s})}, \qquad (3)$$

where $T_0$ is the transparency coefficient of the waveguide end, $E_{h,s}^{ext}(t)$ are the external fields launched into the resonator (below, we will consider a more general case when $E_{h,s}^0$ are time-dependent), $\delta_{h,s} = 2k_{h,s}(L+x)$ is the phase of the round-trip pass through the resonator; $x$ is the coordinate of the microcantilever; and $k_{h,s}$ are the wave numbers.

The motion of the microcantilever is described by the equation for a harmonic oscillator under perturbed by radiation pressure force and thermal noise:

$$\ddot{x} + \Gamma\dot{x} + \omega_0^2 x = A|E_h|^2 + A\left[E_h^* E_s \exp(-i\omega_0 t) + c.c.\right] + \frac{F}{m}, \qquad (4)$$

where $A = \dfrac{S}{4\pi m}$, $S$ is the area of microcantilever surface, $m$ is the microcantilever mass; and $F$ is the thermal noise force. We assume that the difference between the heterodyne frequency and the signal currier frequency is equal to the cantilever fundamental frequency $\omega_0 = \omega_h - \omega_s$. To obtain an analytical solution, we use the Fourier transform equations corresponding to Eqs. (2,4)

$$i\omega E_{h,s}(\omega) = -\frac{1}{2\tau_p}\left(E_{h,s}(\omega) - E_{h,s}^0(\omega)\right), \qquad (5a)$$

$$\left(\omega_0^2 - \omega_1^2 - i\omega_1\Gamma\right)x(\omega_1) = A\int_{-\infty}^{+\infty}d\omega\left[E_h(\omega)E_s^*(\omega+\omega_0+\omega_1) + E_h^*(\omega)E_s(\omega+\omega_0-\omega_1)\right] + F(\omega_1)/m. \qquad (5b)$$

Eq. (5b) does not include the non-resonant term $A|E_h|^2$. It is known[9] that this term leads to a steady state shift of the cantilever amplitude to a new point of equilibrium, and it changes the frequency and damping rate of the cantilever oscillations. The analysis of influence of this term on the sensitivity of the detector is performed in Section III. We assume that the signal is the stationary broadband field emitted by a thermal object. The $\delta$-correlated signal is given by the expression

$$\frac{A}{2}\left\langle E_s^{ext}(\omega)E_s^{ext*}(\omega')\right\rangle = P_s(\omega)\delta(\omega-\omega'), \qquad (6)$$

where $P_s(\omega)$ is the spectral density of the signal power incident on the resonator. The heterodyne field is described as a stationary field with the Lorentzian spectral distribution,

$$\frac{A}{2}\left\langle E_h^{ext}(\omega)E_h^{ext*}(\omega')\right\rangle = P_h\frac{\Gamma_h}{\pi(\omega^2+\Gamma_h^2)}\delta(\omega-\omega'), \qquad (7)$$

where $P_h$ is the external heterodyne power. For the spectral components of the thermal noise force, we have

$$\left\langle F(\omega)F^*(\omega')\right\rangle = \delta(\omega-\omega')\frac{k_B T K \Gamma}{\omega_0^2 \pi}, \qquad (8)$$

where $K = m\omega_0^2$ is the spring constant of a microcantilever, $k_B$ is the Boltzmann constant; and $T$ is the temperature.

Below we calculate the minimal measurable spectral irradiance (MMSI) using the equality,

$$\langle x_s^2(t)\rangle = \langle x_T^2(t)\rangle, \qquad (9)$$

where $\langle x_s^2(t)\rangle$ is the mean square amplitude of oscillations induced by the radiation pressure force neglecting thermal noise; and $\langle x_T^2(t)\rangle = k_B T/K$ is the thermal noise mean square amplitude. Note that the system of equations for the harmonic oscillator and the field in the optical cavity (Eqs. 5a,b) is a nonlinear problem due to the nonlinear dependence of the field amplitude in the resonator on the microcantilever coordinate given by Eq. (3). We consider a linear approximation for the solution of Eqs. (5a,b). We assume that for realistic values of parameters the oscillation amplitude of the cantilever is much smaller than the region of dispersion of the optical resonator. Our numerical simulations, presented below, confirm this assumption. The equilibrium position of the microcantilever can be chosen to provide the maximal field enhancement inside the resonator. In this case Eq. (3) gives

$$E_{h,s}^0(\omega) = \frac{E_{h,s}^{ext}(\omega)T_0}{1-R}. \qquad (10)$$

We assume that the line width of the heterodyne source, $\Gamma_h$, the microcantilever resonance frequency, $\omega_0$, and the damping rate, $\Gamma$, are much less than the bandwidth of the optical resonator, $\Gamma_h, \omega_0, \Gamma \ll \tau_p^{-1}$. The frequency deviation of the signal spectral components from the heterodyne carrier frequency is of the order of $\omega_0$. (See Eq. (12) below.) These frequency differences are negligibly small compared with the resonator line width. Thus Eq. (10) could be satisfied simultaneously for the heterodyne field and for the signal field.

To calculate the mean square amplitude $\langle x_s^2(t)\rangle$, we use the solution of Eqs. (5a,b)

$$x_s(t) = A \int_{-\infty}^{\infty} d\omega_1 \int_{-\infty}^{\infty} d\omega \frac{\exp(-i\omega_1 t)(E_h(\omega)E_s^*(\omega+\omega_0+\omega_1) + E_h^*(\omega)E_s(\omega+\omega_0-\omega_1))}{\omega_0^2 - \omega_1^2 - i\omega_1 \Gamma}. \tag{11}$$

Taking into account Eqs. (6,7,10,11), we obtain the following expression for the mean square amplitude:

$$\langle x_s^2(t) \rangle = B \int_{-\infty}^{\infty} d\omega_1 \int_{-\infty}^{\infty} d\omega \frac{P_s(\omega+\omega_0-\omega_1)}{(\omega^2 + \Gamma_h^2)(\omega^2 + \tau_p^{-2})[(\omega+\omega_0-\omega_1)^2 + \tau_p^{-2}][(\omega_1^2 - \omega_0^2)^2 + \omega_1^2 \Gamma^2]}, \tag{12}$$

where $B = \dfrac{T_0^4 P_h \Gamma_h}{4\pi m^2 c^2 (1-R)^4 \tau_p^4}$.

The integrals in Eq. (12) are calculated assuming that (i) $\Gamma_h \ll \tau_p^{-1}$ and (ii) the signal spectral band is much broader than the line widths of the heterodyne laser and the optical resonator. Consequently, when calculating the integral over $\omega_1$ in Eq. (13), we can take into account only residuals of the denominator. Performing these straightforward calculations, we obtain

$$\langle x_s^2(t) \rangle = \frac{4\pi P_h \omega_0^2 T_0^4 [P_s(0) + P_s(2\omega_0)]}{c^2 K^2 \Gamma (1-R)^4}. \tag{13}$$

According to Eq. (12) the microcantilever senses the sum of two spectral components corresponding to the frequencies $\omega_s = \omega_h - \omega_0$ and $\omega_s = \omega_h + \omega_0$. As assumed above, the signal bandwidth is much larger than the cantilever frequency, $\omega_0$. In our later considerations we will assume $P_s(0) = P_s(2\omega_0)$. Combining Eqs. (9) and (13) we obtain for MMSI

$$I_{\min} = \frac{k_B T \rho d c^2 \Gamma (1-R)^4}{8\pi P_h T_0^4}, \tag{14}$$

where $\rho$ is the density of the cantilever material, $d$ is the thickness of the cantilever. We estimate the MMSI for the values of parameters typical for two types of cantilevers. First is the thin and soft cantilever usually used for atomic force microscopy. Second is the micro-mirror with a very high coefficient of reflection utilized in the laser cooling experiments[10,11].

*Example 1.* In the case of thin and soft cantilever, the typical values of parameters are: $\rho = 2.33 \ kg/m^3$, $d = 60 \ nm$, $\Gamma = 2\pi \times 10^2 \ s^{-1}$, $\Gamma_h = 2\pi \times 10^8 \ s^{-1}$, and $\lambda = 10 \ \mu m$, $T = 300 \ K$, $R = 0.95$, $T_0^2 = 0.1$, and $P_h = 10^{-3} \ W$. For these values of the parameters, Eq. (14) gives $MMSI = 5.1 \times 10^{-6} \ W/m^2 Hz$. The response time is $\tau = 2\pi \ \Gamma^{-1} = 10 \ ms$. For the chosen reflection coefficient, $R$, and the length of the resonator, $L = 15 \ \mu m$, the damping rate is, $2\tau_p^{-1} = 5 \times 10^{11} \ s^{-1}$. Consequently our assumption that $2\tau_p^{-1} \gg \Gamma_h$, $\omega_0$, $\Gamma$, is fulfilled. The dispersion length of the optical resonator is $l = (1-R)\lambda/(4\pi\sqrt{R}) = 40 \ nm$. The average amplitude of the thermal oscillations is $\sqrt{\langle x_T^2 \rangle} = \sqrt{k_B T/K} = 0.65 \ nm$. Thus, the condition of the validity of the linear approximation, $l \gg \sqrt{\langle x_T^2 \rangle}$, is also fulfilled.

*Example 2.* In the case of the high reflection coefficient[10,11] mirror with an area $S = 520 \times 120 \ \mu m^2$, the parameters are: $\rho = 2.67 \ kg/m^3$, $d = 2.4 \ \mu m$, $\Gamma = 2\pi \times 13 \ s^{-1}$, and $T = 300 \ K$, $R = 0.998$, $T_0^2 = 2 \times 10^{-3}$, $P_h = 10^{-1} \ W$. In this case, $MMSI = 1.95 \times 10^{-9} \ W/m^2 Hz$. The response time is $\tau = 2\pi \ \Gamma^{-1} = 77 \ ms$. Comparison of these two cases shows that most critical parameter is the coefficient of reflection, $R$. The value of the reflectance obtained for micromechanical Bragg mirrors[10,11] is not highest for this technology. As authors[11] note, the Bragg mirror technology can provide the reflection coefficient even higher than 0.999999. Mirrors with reflectance 0.9999 for infrared radiation are commercially available. The adaptation of this technology for micromechanical mirrors leads to tremendous progress in the sensitivity. For example, using the value of a reflection coefficient $R = 0.9999$, and the values of others parameters used in Example 2, we obtain $MMSI = 4.87 \times 10^{-12} \ W/m^2 Hz$. The spectral resolution of the microcantilever based spectrometer described above is defined by the heterodyne laser line width. For comparison of the sensitivity of the proposed spectrometer with a standard IR-spectrometer with uncooled detector, consider a numerical example. If the heterodyne laser has the line width, $\Delta\nu = 30 \ MHz$, then the minimal measurable intensity is $MMSI \times \Delta\nu = 1.5 \times 10^{-4} \ W/m^2$. The sensitivity of a thermal radiation detector, such as a bolometer, is characterized by the noise equivalent temperature difference (NETD), which typically is $50\ mK$ for a room temperature source emitting in the spectral interval, $8-14\ \mu m$. The corresponding minimal measurable intensity is $0.13\ W/m^2$. Thus, the microcantilever spectrometer has a sensitivity that is three orders of magnitude better.

In our estimations we use a rather high quality factor of the microcantilever oscillator, $\omega_0/\Gamma = 10^3 - 2\times 10^4$. This quality factor is typical for microcantilevers placed in a vacuum chamber. The vacuum technology is well developed and is commonly used in microcantilever based thermal imagers[4].

### III. Thermal radiation detector

Thermal imaging usually exploits broad band infrared radiation corresponding to the window of atmospheric transparency $8-14\ \mu m$. For these applications we consider a scheme in which IR radiation intensity is modulated by the optical modulator (see Fig. 2) at a frequency equal to the resonance cantilever frequency, $\omega_0$. In order to fulfill the resonance condition for each spectral amplitude of the radiation field, a certain spectral amplitude must be sent into the resonator at the proper angle. This could be obtained, for example, using a volume holographic grating (see Fig. 2). The intensity modulation coefficient can be presented in the form

$$r(t) = \int_{-\infty}^{\infty} d\omega\, r(\omega) \exp i(\omega_0 - \omega)t + c.c., \qquad (15)$$

Where the function $r(\omega)$ is

$$\langle r(\omega) r^*(\omega') \rangle = r^2 \frac{\Gamma_M}{\pi(\omega^2 + \Gamma_M^2)} \delta(\omega - \omega'), \qquad (16)$$

and $2\Gamma_M$ can be considered as a measurement bandwidth. Then, the average mean square amplitude of thermal oscillation is $\sqrt{\langle x_T^2 \rangle} = \sqrt{k_B T 2\Gamma_M / K\pi\Gamma}$.

Straightforward calculations, similar to those performed in Section III, give the estimate for a minimal detection intensity

$$I_s = r^{-1} \left[ \frac{2 k_B T \rho d c^2 \Gamma_M \left( \frac{\Gamma}{2} + \Gamma_M \right)(1-R)^2}{\pi S} \right]^{1/2}. \qquad (17)$$

Choosing $\Gamma_M = \Gamma = 13 Hz, R = 0.99996, r = 1, S = 520 \times 120 \mu m^2, \rho = 2.67 kgm^{-3}, d = 2.4 \mu m$ we have for the minimal detected intensity, $I_s = 2 \times 10^{-2} \ W/m^2$. To calculate the NETD we have to use the slope of the black body radiation within the spectral band 8-14 $\mu m$: $(dP/dT)_{\lambda_1 - \lambda_2} = 2.62 \ Wm^{-2} K^{-1}$. Then, we obtain NETD = $I_s / (dP/dT_{\lambda_1 - \lambda_2})$=7.6 mK, with the response time $2\pi/\Gamma = 33 ms$. This value of NETD is an order of magnitude less than conventional uncooled bolometers usually have, and it is comparable with a theoretical limit, 9.2 mK, for the thermo-mechanical detectors[4]. Note that the radiation pressure measurements are less affected by temperature fluctuations compared with thermo-mechanical detectors.

## IV. Numerical simulations

For numerical simulations we use Eqs. (2,4). The thermal noise and the signal are simulated by a standard random number generator, which produces a random sequence of pulses. The time duration of these pulses, $\Delta t$, is much shorter than the oscillation period of the microcantilever, $T = 2\pi/\omega_0 = 40 \Delta t$. The probability of the pulse amplitude is uniform in the interval $[-a_{th,s}, a_{th,s}]$. To avoid undesirable correlations between the signal and thermal noise processes, we use two different generators of random pulses. We find that for nonlinear oscillations, when the amplitude of thermal oscillations, $x_T$, is comparable with the optical resonator dispersion length, $l$, the nonlinearity leads to a significant decrease in the signal-to-noise ratio. Thus, the nonlinearity is an undesirable feature. The values of parameters, chosen above in Section II, provide a linear regime of microcantilever oscillations. To control the linearity, we calculate the mean square amplitude for the thermal noise $\langle x_T^2(t) \rangle$ only (with a zero signal), then calculate the mean square signal amplitude $\langle x_s^2(t) \rangle$ without the thermal noise, and calculate the mean square amplitude $\langle x^2(t) \rangle$ for conditions in which both noises exist. In the linear case, the last sum is equal to the sum of mean square amplitudes calculated separately:

$$\langle x^2(t) \rangle = \langle x_T^2(t) \rangle + \langle x_s^2(t) \rangle.$$

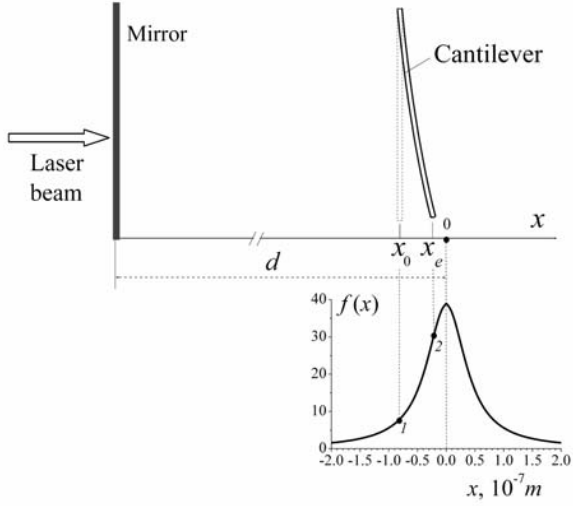

Fig. 3. Schematic diagram for the initial and equilibrium positions of the microcantilever, $f(x) = |T_0/[1 - R\exp(\delta_{h,s})]|^2$ (see Eq. (3)).

As shown in the Section II, the highest sensitivity is obtained when the optical resonator is tuned to exact resonance with the both heterodyne and signal fields. However, the non-resonant term in Eq. (4) induces a shift of the equilibrium position of the amplitude of oscillations. Nevertheless, the optimal position can be obtained by introducing an initial off-set of the position of the cantilever relative to the resonance. (See Fig. 3.) This could be achieved, for example, by using a feedback loop which measures the level of the heterodyne intensity after passing the resonator and feeds it back to a piezo-ceramic substrate which shifts the initial position of the cantilever. The dependence of the optimal equilibrium position of the cantilever, $x_e$, on the heterodyne power for different initial deviation, $x_0$, is presented in Fig. 4. These dependences demonstrate the well-known[12] bi-stable type of curves. To avoid undesirable instability, a practically acceptable region must be chosen outside of the bi-stable region.

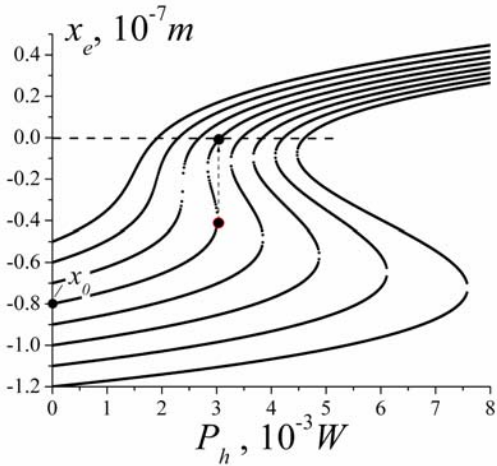

Fig. 4. Dependence of a microcantilever equilibrium position $x_e$ on the power of the heterodyne laser.

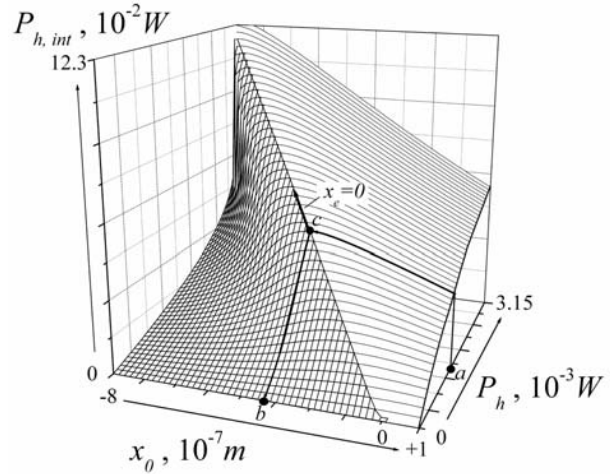

Fig. 5. Dependence of the intro-cavity heterodyne power on the initial position of the microcantilever and external heterodyne power.

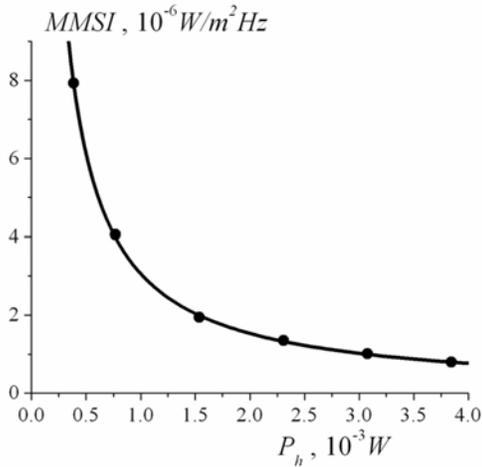

Fig. 6. Dependence of MMSI on the heterodyne power.

A 3-D representation of possible values of initial deviations and external and internal values of the heterodyne power is presented in Fig. 5. The curve along the "ridge" corresponds to the maximally enhancement values. The dependence of the MMSI on the heterodyne external power, $P_h$, for the optimal initial deviation for each value, $P_h$, is presented in Fig. 6. As can seen, the value $P_h = 10^{-3}$ $W$ corresponds to $MMSI = 3.5 \times 10^{-6}$ $W/m^2 Hz$, which is in qualitative agreement with the analytical estimate given in Section 2 (Example 1) for the same value of the heterodyne power, $MMSI = 5.1 \times 10^{-6}$ $W/m^2 Hz$.

## IV. Conclusion

We have described a far infrared (terahertz), uncooled detector based on a microcantilever as a radiation pressure sensor, which is expected to have a significantly higher sensitivity than existing uncooled detectors in the far infrared-terahertz spectral region. The significant enhancement of sensitivity is due the combination non-absorption detection method and high quality optical microcavity.